\documentclass[fleqn,usenatbib]{mnras}
\usepackage{newtxtext,newtxmath}
\usepackage[T1]{fontenc}

\usepackage{graphicx}	
\usepackage{amsmath}

\title[Constraints on Electron Acceleration from GRB Radio Peaks]{Constraints on Electron Acceleration in Gamma-Ray Bursts Afterglows from Radio Peaks}

\author[]{
Ruby A. Duncan,$^{1}$\thanks{E-mail: rubyduncan@gwu.edu}
Alexander J. van der Horst$^{1}$, Paz Beniamini$^{2,3}$
\\
$^{1}$Department of Physics, The George Washington University, 725 21st Street NW, Washington, DC 20052, USA\\
$^{2}$Department of Natural Science, The Open University of Israel, P.O. Box 808, Ra'anana 4353701, Israel\\
$^{3}$Astrophysics Research Center of the Open University (ARCO), The Open University of Israel, P.O. Box 808, Ra'anana 4353701, Israel
}

\date{1 November 2022}

\pubyear{2022}

\begin{document}

\label{firstpage}
\pagerange{\pageref{firstpage}--\pageref{lastpage}}

\maketitle

\begin{abstract}
Studies of gamma-ray bursts (GRBs) and their multi-wavelength afterglows have led to insights in electron acceleration and emission properties from relativistic, high-energy astrophysical sources. Broadband modeling across the electromagnetic spectrum has been the primary means of investigating the physics behind these sources, although independent diagnostic tools have been developed to inform and corroborate assumptions made in particle acceleration simulations and broadband studies. We present a methodology to constrain three physical parameters related to electron acceleration in GRB blast waves: the fraction of shock energy in electrons, $\epsilon_e$; the fraction of electrons that gets accelerated into a power-law distribution of energies, $\xi_e$; and the minimum Lorentz factor of the accelerated electrons, $\gamma_m$. These parameters are constrained by observations of the peaks in radio afterglow light curves and spectral energy distributions. From a sample of 49 radio afterglows, we are able to find narrow distributions for these parameters, hinting at possible universality of the blast wave microphysics, although observational bias could play a role in this. Using radio peaks and considerations related to the prompt gamma-ray emission efficiency, we constrain the allowed parameter ranges for both $\epsilon_e$ and $\xi_e$ to within about one order of magnitude, $0.01\lesssim\epsilon_e\lesssim0.2$ and $0.1\lesssim\xi_e\lesssim1$. Such stringent constraints are inaccessible for $\xi_e$ from broadband studies due to model degeneracies.
\end{abstract}


\section{Introduction}

The multi-wavelength afterglow of gamma-ray bursts \citep[GRBs;][]{vanParadijs_1997, Costa_1997, Frail_1997} allows for detailed studies of the physics behind these highly energetic cosmic explosions. Observations across the electromagnetic spectrum can be used to model their extreme, collimated outflows. From the afterglow observations one can derive physical parameters characterizing the energetics of the explosion, the circumburst medium, and the microphysics of the observed radiation \citep{Sari_1998, Wijers_1999}. The latter includes parameters describing the acceleration of the emitting particles, predominantly electrons, and the magnetic fields in which these electrons are accelerated and radiate. 

From GRB afterglow observations, a spectral energy distribution (SED) can be constructed over a broad frequency range, from radio wavelengths to gamma-ray energies \citep[e.g.,][]{Perley_2014, Magic_2019}. The dominant radiation process is synchrotron emission, with inverse Compton effects important at X-ray and higher energies \citep{Sari_2001}. The rapidly evolving SED results in varying light curves at all observing frequencies, from which one can study the different stages of the afterglow evolution. Observations in the radio regime are particularly useful in constructing the SED, as they can determine the evolution of the spectral peak and the synchrotron self-absorption frequency \citep{Frail_1997, Wijers_1999}, two observables necessary to constrain the physical parameters of the GRB jet and its environment.

Multi-wavelength studies of afterglows are predominantly performed within the relativistic blast wave or fireball model \citep{Rees_1992}. In this framework, a relativistic shock at the front of the jet accelerates electrons that produce the observed radiation. Broadband modeling of the afterglow observations have been carried out with (semi-)analytical approximations of the blast wave evolution \citep[e.g.,][]{Chevalier_1999, Granot_2002, Panaitescu_2002, Yost_2003}, but more recently this has also been done by fitting hydrodynamical simulations directly to the data \citep{vanEerten_2012}. Broadband modeling of a well-sampled afterglow can result in tight constraints on all the physical parameters. While these methods have proven to be powerful for studying the physics of the outflow and its environment, the results reported show wide ranges for each of the parameter values \citep{Cenko_2011}, as well as conflicting results across studies. Even in the cases of modeling a singular GRB afterglow, different methods and modeling codes can produce widely varying values for the physical parameters \citep{Granot_2014}. This makes it, for instance, difficult to establish the widths of distributions for these parameters. Several studies have suggested the possibility of universal values for some of the parameters \citep[e.g.,][]{Kirk_2000,Achterberg_2001}, especially the micro-physical ones, but the current broadband modeling studies do not allow for pinning this down.

To support the estimation of parameters through broadband modeling, there have been a few studies focused on developing alternative means for parameter estimation, in particular using only a few observables to constrain parameter values independently. \citet{Nava_2014} derived narrow distributions for two of the physical parameters, the fraction of shock energy in electrons and the efficiency of the prompt radiation mechanism, through normalized light curves at high-energy gamma rays. \citet{Beniamini_2015} used a combination of the high-energy gamma-ray and X-ray emission to find a narrow distribution for the efficiency of the prompt emission as well. 

The work presented in this paper focuses on another diagnostic tool: using the peaks in radio afterglows to constrain physical parameters that describe the physics of electrons accelerated by the blast wave. \citet{Beniamini_2017} used the peaks in radio light curves to constrain the fraction of shock energy in electrons, $\epsilon_e$. They concluded that $\epsilon_e$ is relatively narrowly distributed around $0.13-0.15$, with a width in log space of $0.26-0.31$. We expand upon this work by increasing the sample size and taking into account the peaks in broadband radio SEDs as well, thus further constraining $\epsilon_e$, and placing constraints on two more physical parameters: the minimum Lorentz factor of the accelerated electrons, $\gamma_m$, and the fraction of electrons accelerated by the shock to a power-law distribution, $\xi_e$. 

Our sample of radio light curves and SEDs is outlined in Section \ref{sample}, and the methodology we follow is described in Section \ref{sec:method}. Our results are presented in Section \ref{results}, and in Section \ref{discuss} we discuss our findings, with our conclusions given in Section \ref{conclusions}. We adopt the cosmological parameters $H_0$ = 69.6 km/s/Mpc, $\Omega_{\rm{M}}$ = 0.286, and $\Omega_\Lambda$ = 0.714.

\section{Radio Sample} \label{sample}

The sample used in this work consists of GRBs with radio afterglow observations that provide well-sampled light curves or SEDs needed to determine a peak. \citet{Beniamini_2017} used a sample of 36 afterglows compiled from the \citet{Chandra_Frail_2012} catalog of GRB radio observations from 1997 to 2011, with a majority of the data taken with the Very Large Array (VLA) at 8.5~GHz. For the new sample, we use bursts observed from 2010 to 2019, from different radio observatories: the upgraded VLA, Westerbork Synthesis Radio Telescope (WSRT), Arcminute Microkelvin Imager Large Array (AMI-LA), Combined Array for Research in Millimeter-wave Astronomy (CARMA), and Karoo Array Telescope (MeerKAT). 

The sample has GRBs for which radio peaks could be determined in either light curves or SEDs, with measurements ranging from from 1.3 to 15.7 GHz and 4 to 31 days. All GRBs in our sample have a spectroscopic redshift and a well-determined isotropic equivalent gamma-ray energy, $\text{E}_{\gamma, \text{iso}}$. This results in a sample size of 13 afterglows (hereafter referred to as Sample 2), which we add to the \citet{Beniamini_2017} data set (Sample 1), totaling 49 GRBs. For well-sampled afterglows, we were able to identify peaks in more than one light curve or SED. In those cases, we used the best-constrained peak in our analysis. Results for the other peaks are included in Table \ref{tab:peak_params} and Table \ref{tab:psi_chi}, denoted in italics. They have not been included in the statistical analysis on the physical parameters derived from the radio peaks, nor in the results discussed in Section~\ref{psi_results}. 

The 36 afterglows from Sample 1 have a peak flux and time from the \citet{Chandra_Frail_2012} catalog. In order to obtain the peak parameters for Sample 2, we fit a smoothly broken power-law model to each well-defined peak, with a fairly sharp smoothness parameter ($s=5$). The free parameters in the fitting routine are the peak flux $F_{\nu_p}$, the peak time $t_p$ (for light curves) or peak frequency $\nu_p$ (for SEDs), and the pre- and post-break power-law slopes. The fit results for every GRB are listed in Table \ref{tab:peak_params}. 

For Sample 1, 21 of the GRBs had a jet-break time derived from achromatic breaks in optical and/or X-ray light curves, obtained from \citet{Chandra_Frail_2012}. 
10 of the 13 GRBs in Sample 2 have jet break times reported in the literature, which we give in Table \ref{tab:peak_params}.

\begin{table*}
\caption{Radio peak and other relevant parameters for Sample 2, from light curve and SED modeling of observations from the Very Large Array (VLA), Westerbork Synthesis Radio Telescope (WSRT), Arcminute Microkelvin Imager Large Array (AMI-LA), Combined Array for Research in Millimeter-wave Astronomy (CARMA), and Karoo Array Telescope (MeerKAT).
[1] \citet{Laskar_2015}; [2] \citet{Antonelli_2010}; [3] \citet{Marshall_2011}; [4] \citet{deUgarte_2018}; [5] \citet{Moin_2013}; [6] \citet{Chornock_2010}; [7] \citet{vanderHorst_2015}; [8] \citet{Zauderer_2013}; [9] \citet{Tello_2012}; [10] \citet{Laskar2018C}; [11] \citet{Tanvir_2014}; [12] \citet{Higgins_2019}; [13] \citet{Anderson_2018}; [14] \citet{Cucchiara_2015}; [15] \citet{Perley_2014}; [16] \citet{deUgarte_2015}; [17] \citet{Golenetskii_201}; [18] \citet{Laskar2016}; [19] \citet{Tanvir_2016}; [20] \citet{Alexander_2017}; [21] \citet{Xu_2016}; [22] \citet{Zhang_2018}; [23] \citet{Laskar_2018a}; [24] \citet{Tanvir_2016B}; [25] \citet{Alexander_2019}; [26] \citet{Kankare_2017}; [28] \citet{Valeev_2019}; [29] \citet{Rhodes_2020}}

\begin{tabular}{llllllllll} \hline \hline
GRB     & $F_{\nu_p}$  & $\nu_p$     & $t$            & $d_{28}$ & $z$                   & $E_{\gamma, \text{iso}}$ & $t_j$           & Telescope        & Ref. \\
        & (mJy)             & (Ghz)             & (d)               & $(10^{28} \text{cm})$         &                       & $(10^{53} \text{ergs})$       & (d)               &               &  \\ \hline
\emph{100418A} & \emph{0.58 $\pm$ 0.06}   & \emph{4.8}            & \emph{38 $\pm$ 21}   & \emph{1.15}                        & \emph{0.624} & \emph{0.0099}                                         & \emph{17}        & \emph{VLA, WSRT}  & \emph{1,2,3,4,5} \\
100418A & 1.42 $\pm$ 0.08   & 8.46           & 52.8 $\pm$ 6.1  & 1.15                         & 0.624 & 0.0099                                         & 17        & VLA        & 1,2,3,5   \\
100901A & 0.32 $\pm$ 0.02   & 4.8            & 9.4 $\pm$ 1.7  & 3.16                         & 1.41  & 0.8                                            & 0.96      & VLA, WSRT  & 1,6       \\
111215A & 0.84 $\pm$ 0.13   & 4.8            & 24.3 $\pm$ 6.5  & 5.06                         & 2.01    & 0.45                                           & $>31$     & VLA, WSRT  & 7,8       \\
120326A & 0.83 $\pm$ 0.08   & 28.7 $\pm$ 7.3 & 15.4            & 4.28                         & 1.80  & 0.32                                           & 1.5       & VLA, CARMA & 1,9       \\
\emph{120326A} & \emph{0.79 $\pm$ 0.10}    & \emph{10.3 $\pm$ 1.6} & \emph{31.3}            & \emph{4.28}                         & \emph{1.80}  & \emph{0.32}                                           & \emph{1.5}       & \emph{VLA}       & \emph{1,9}       \\
140311A & 0.42 $\pm$ 0.06   & 23.0 $\pm$ 4.0   & 4.5             & 13.3                         & 4.59   & 2.7                                            & 0.6       & VLA        & 10,11     \\
140713A & 1.67 $\pm$ 0.14   & 15.7           & 11.1 $\pm$ 0.8  & 1.90                          & 0.935  & 0.017                                          & 30        & AMI-LA       & 12,13     \\
141121A & 0.22 $\pm$ 0.03   & 15.7           & 16.4 $\pm$ 3.4  & 3.34                         & 1.47  & 0.80                                            & 3         & AMI-LA, VLA   & 13,14,15  \\
150413A & 0.25 $\pm$ 0.06   & 15.7           & 4.7 $\pm$ 0.4 & 8.45                         & 3.14   & 6.5                                            & ...       & AMI-LA       & 13,16,17  \\
160509A & 1.39 $\pm$ 0.09   & 8.1 $\pm$ 0.6 & 4.1             & 2.51                         & 1.17   & 5.76                                           & 6         & VLA        & 18,19  \\
\emph{160509A} & \emph{0.59 $\pm$ 0.02}   & \emph{5.0}            & \emph{2.3 $\pm$ 0.2}  & \emph{2.51}                         & \emph{1.17}   & \emph{5.76}                                           & \emph{6}         & \emph{VLA}        & \emph{18,19}  \\
\emph{160625B} & \emph{0.61 $\pm$ 0.19}   & \emph{2.1 $\pm$ 0.8} & \emph{12.5}            & \emph{3.15}                         & \emph{1.41}  & \emph{30}                                             & \emph{25}        & \emph{VLA}        & \emph{20,21,22}  \\
160625B & 0.59 $\pm$ 0.07    & 1.9 $\pm$ 0.4 & 22.5            & 3.15                         & 1.41  & 30                                             & 25        & VLA        & 20,21,22  \\
161219B & 3.90 $\pm$ 0.28    & 22.8 $\pm$ 1.1   & 1.52            & 0.13                         & 0.148 & 0.0018                                         & 32        & VLA        & 23,24,25  \\
171010A & 1.35 $\pm$ 0.02   & 14.2 $\pm$ 0.8 & 9.66            & 0.53                         & 0.33   & 2.2                                            & ...       & VLA        & 26,27     \\
190829A & 1.67 $\pm$ 0.33   & 1.3            & 12.9 $\pm$ 3.8  & 0.011                         & 0.0785 & 0.0030                                        & ...       & MeerKAT    & 28,29    
\end{tabular}
\label{tab:peak_params}
\end{table*}

\section{Methodology} \label{sec:method}

The afterglow of GRBs can be described by an expanding, relativistic shell or blast wave propagating outward through an external medium \citep{Rees_1992}. The broadband synchrotron radiation we observe is emitted by electrons accelerated by the shock interacting with the ambient medium, producing the afterglow that can be seen from X-ray to radio wavelengths on timescales of seconds to years \citep[e.g.,][]{Sari_1998,Wijers_1999}.

The dynamics of the blast wave is characterized by its isotropic equivalent kinetic energy, $E_{iso}$, which is driving the shock outward through the ambient medium, where the density of the medium, $n$, also plays a role in the blast wave evolution. The ambient medium is typically assumed to be homogeneous or structured as a stellar wind, with the density dropping off as one over the distance from the central engine squared \citep{Chevalier_1999}. The electrons behind the shock are accelerated to a power-law distribution of energies, with power-law index $p$; and with $\epsilon_e$ describing the ratio of shock energy in electrons to the total energy density. Similarly, $\epsilon_B$ describes the ratio of energy density in the magnetic field to the total energy. 

These micro- and macrophysical parameters can be determined from several observables that describe the synchrotron emission spectrum. The latter is typically characterized as a series of connected power-laws, with three characteristic frequencies connecting each regime: the peak frequency $\nu_m$, the cooling frequency, $\nu_c$, and the synchrotron self-absorption frequency, $\nu_a$. These observable parameters, as well as the peak flux, evolve with time and determine the shape of the spectrum at a given time \citep[e.g.,][]{Granot_2002}. $\nu_c$ is typically found in the X-ray or optical bands, while $\nu_m$ is found to evolve from the optical to the radio regime (as a power law in time with slope -3/2 or even faster than that), and $\nu_a$ is usually detected in the radio from the onset of a GRB. As these three characteristic frequencies evolve over time, they will cross through observing frequencies, creating light curve breaks or peaks. The latter can also occur across the electromagnetic spectrum at the same time, which is caused by dynamical rather than spectral effects. Examples of this are the jet-break time $t_j$, which is related to the opening angle of the jet, or the time at which the blast wave becomes sub- or non-relativistic.

In this work, we assume that the observed radio peaks are due to $\nu_m$. While there are scenarios in which $\nu_m$ is below $\nu_a$, resulting in the peak of the SED being $\nu_a$ instead of $\nu_m$, for typical physical parameters this occurs at late times (months after the GRB onset) and is mostly observed at low radio frequencies \citep[e.g.][]{vanderHorst_2008,Beniamini_2017}. We confirmed this by checking that the slopes in both light curves and SEDs indeed match the theoretical expectations for $\nu_m$ being the peak \citep{Granot_2002}. Furthermore, if the radio peak is caused by $\nu_a$ instead of $\nu_m$, the derived $\Psi$ parameter (see Section~\ref{psi_proxy}) would deviate strongly from the typical range and could have an unphysical value (see also Section~\ref{evolution}). The latter is also true if the observed radio peak is caused by the reverse shock instead of the forward shock \citep[see also][]{Beniamini_2017}.

A fraction $\xi_e$ of the total number of electrons is accelerated into a power-law distribution of electron energies or Lorentz factors, with a minimum Lorentz factor $\gamma_m$, and emit synchrotron radiation. In many broadband modeling studies, $\epsilon_e$ is assumed to describe the entire population of electrons, with $\xi_e$ equal to 1. This assumption is commonly made since $\xi_e$ is degenerate with the energy, density, and $\epsilon_e$ and $\epsilon_B$ in standard broadband modeling \citep{Eichler_Waxman_2005}; and broadband modeling efforts including a variable $\xi_e$ result in poor constraints on its value \citep[e.g.,][]{cunningham_2021}. However, it has been shown that $\xi_e\ll1$ is necessary for prompt emission models that are based on synchrotron emission from a kinetic dominated outflow, such as internal shocks \citep{Daigne_1998,Bosnjak_2009,Beniamini_2013}. In the following subsections we show that constraints on $\xi_e$ at the forward shock can be obtained from the peaks in radio afterglows, with a less stringent degeneracy that hampers such constraints from broadband modeling.

\subsection{$\Psi$ as A Proxy for $\epsilon_e$ and $\xi_e$} \label{psi_proxy}

\citet{Beniamini_2017} presented a framework in which they defined a parameter $\Psi$ as a proxy for $\epsilon_e$, which can be derived from observational properties of the peaks in radio light curves. 
Based on equations for the peak flux and peak frequency as a function of time and the aforementioned physical parameters \citep{Granot_2002}, \citet{Beniamini_2017} derived equations for the parameter $\Psi$ from the ratio of the peak flux and peak frequency. 
We have modified the equations for $\Psi$ by including the parameter $\xi_e$, which was set equal to 1 in \citet{Beniamini_2017}, to investigate the relationship between $\epsilon_e$ and $\xi_e$.   
Following \citet{Beniamini_2017}, we show the equations for a homogeneous circumburst medium, such as expected for the interstellar medium (ISM), as well as a stellar wind (Wind):

\begin{equation}\label{psi_ism}
\begin{split}
    \Psi_{\text{ISM}} &= \left(\frac{261.4~(1+z)^{1/2}~\nu_p~t_p^{3/2}~E_{\gamma,iso,53}^{1/2}}{10^{15}~d_{28}^2~F_{\nu_p}~\text{max}(1,t_p/t_j)^{1/2} }\right)^{1/2} \\
    &=  \frac{(p-2)}{0.177~(p-1)} \left(\frac{p-0.67}{p+0.14}\right)^{1/2} \left(\frac{1-\epsilon_{\gamma}}{\epsilon_{\gamma}}\right)^{-1/4} n_0^{-1/4} \epsilon_e~\xi_e ^{-3/2}
\end{split}
\end{equation}

\begin{equation}\label{psi_wind}
\begin{split}
\Psi_{\text{Wind}} &= \left(\frac{249.4~(1+z)~\nu_p~t_p}{10^{15}~d_{28}^2~F_{\nu_p}}\right)^{1/2} \\
&= \frac{(p-2)}{0.277~(p-1)} \left(\frac{p-0.69}{p+0.12}\right)^{1/2} A_*^{-1/2} \epsilon_e~\xi_e ^{-3/2}
\end{split}
\end{equation}

Equations \ref{psi_ism} and \ref{psi_wind} have been structured such that the top line of each equation contains observable parameters and the bottom line contains the physical parameters. 
The observed peak parameters in these equations are the peak frequency $\nu_p$ in GHz, the peak flux $F_{\nu_p}$ in mJy, and the peak time $t_p$ in days. These three are determined from the afterglow light curves or SEDs. For the other observable parameters, the redshift $z$ is obtained from optical spectroscopy of the afterglow (or in some cases the host galaxy), as well as the luminosity distance $d_{28}$ in units of $10^{28}$~cm; $E_{\gamma,\text{iso}, 53}$ is the isotropic equivalent energy of the burst in units of $10^{53}$~erg; and the jet break time $t_j$, in days, is derived from achromatic light curve breaks where possible. 
For the physical parameters, the density $n_0$ in a homogeneous medium is given in cm$^{-3}$, and for a stellar wind the density is characterized by $A_*$ following \citet{Chevalier_1999}; $\epsilon_\gamma$ is the energy efficiency of the prompt emission, with $\epsilon_\gamma=E_{\gamma,iso}/(E_{\gamma,iso}+E_{K,iso})$; and $p$ the aforementioned electron energy distribution power-law index. As can be seen from Equations \ref{psi_ism} and \ref{psi_wind}$, \Psi$ has a weak dependence on most physical parameters except for $\epsilon_e$ and $\xi_e$, which allows us to use $\Psi$ as a proxy for the combination of parameters $\epsilon_e~\xi_e ^{-3/2}$.

\subsection{$\chi$ as A Proxy for $\gamma_m$ and $\xi_e$}

Following similar steps to deriving $\Psi$ as a proxy for $\epsilon_e$ and $\xi_e$, we derive another proxy, $\chi$, for a parameter related to electron acceleration in GRB jets: $\gamma_m$, the minimum Lorentz factor of the electron distribution; and also $\chi$ has a dependence on $\xi_e$. We use the relation between $\gamma_m$ and $\epsilon_e$ as detailed in \citet{Sari_1998}, with the addition of a linear dependence of $\gamma_m$ on $\xi_e$. Adopting standard equations for the shock Lorentz factor and the equations for $\Psi$ given above, we derive $\chi$, which can be estimated from observables, is linear in $\gamma_m$, and weakly dependent on other physical parameters except for $\xi_e$ (with $t_{d}$ the time in days):

\begin{equation} \label{chi_ism}
\begin{split}
\chi_{\text{ism}} & = 266~\Psi_{\text{ism}}~E_{\gamma,\text{iso},53}^{1/8}~(z+1)^{3/8}~t_{d}^{-3/8}\\
 &= \left(\frac{p-0.67}{0.66~(p+0.14)}\right)^{1/2} \left(\frac{\epsilon_{\gamma}}{1-\epsilon_{\gamma}}\right)^{3/8}~n_0^{-1/8}~\gamma_{m}~\xi_e^{-1/2}
\end{split}
\end{equation}

\begin{equation} \label{chi_wind}
\begin{split}
\chi_{\text{wind}} & = 475~\Psi_{\text{wind}}~E_{\gamma,\text{iso},53}^{1/4}~(z+1)^{1/4}~t_{d}^{-1/4} \\
 &= \left(\frac{p-0.69}{0.65(p+0.12)}\right)^{1/2} \left(\frac{\epsilon_{\gamma}}{1-\epsilon_{\gamma}}\right)^{1/4}~A_*^{-1/4}~\gamma_{m}~\xi_e^{-1/2}
\end{split}
\end{equation}

The top line in these equations contains observables and $\Psi$, and the bottom line of each equation contains the physical parameters. One can see that $\chi$ is a proxy for the combination $\gamma_{m}~\xi_e^{-1/2}$.

\section{Results} \label{results}

We have calculated the values for $\Psi$ and $\chi$, and their uncertainties, for all GRBs in Samples 1 and 2. They are listed in Table~\ref{tab:psi_chi} for both the homogeneous (ISM) and stellar wind (Wind) ambient medium structure. Below we discuss the results for our GRB samples. We note that in Table~\ref{tab:peak_params} we show the results for two different peaks in the case of four GRBs in Sample 2, for instance light curves at two different frequencies or SEDs at two different times. The $\Psi$ and $\chi$ values derived for those different peaks are consistent with each other for all four GRBs. In all the sample statistics below we use the peak that is best sampled or defined by the data, and the other peak is indicated in italic font in Tables~\ref{tab:peak_params} and \ref{tab:psi_chi}.

\subsection{Results for $\Psi$} \label{psi_results}

\begin{figure}
    \centering
    \includegraphics[trim=25 10 25 50, clip, width=\columnwidth]{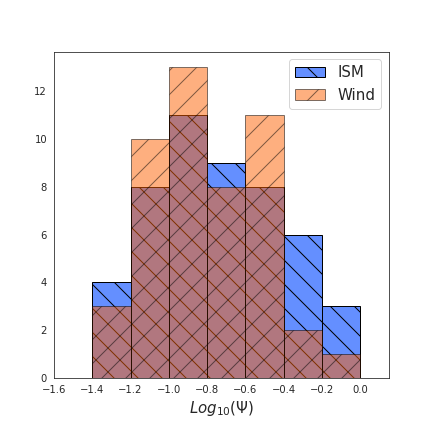}
    \caption{Histogram in log space of $\Psi$ for our GRB sample, for the homogeneous (ISM; blue) and stellar wind (Wind; orange) cases.}
    \label{fig:psi_hist}
\end{figure}

\begin{figure}
    \centering
    \includegraphics[width=\columnwidth]{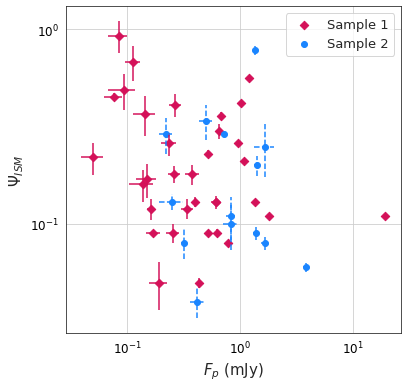}
    \includegraphics[width=\columnwidth]{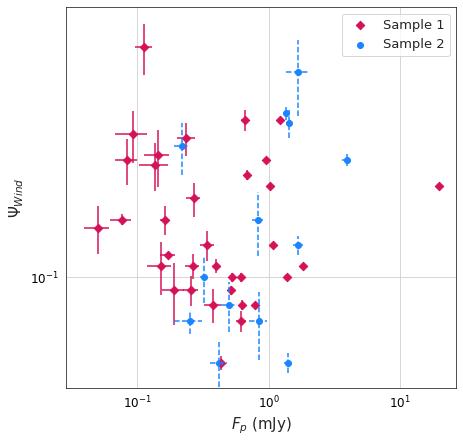}    \caption{$\Psi$ versus peak flux for a homogeneous (top panel) and stellar wind (bottom panel) medium, for the two samples of GRBs presented in this paper.}
    \label{fig:psi_fp}
\end{figure}

Figure~\ref{fig:psi_hist} shows a histogram of $\Psi$ for the ISM (blue) and Wind (orange) circumburst medium cases, for all the GRBs listed in Table~\ref{tab:psi_chi}. The weighted averages for $\Psi$, as well as the widths of the distributions, are consistent between the two samples of GRBs, i.e. those from \citet{Beniamini_2017} and our additional sample. We find that the weighted averages for our total sample (Samples 1 and 2 combined) are $\Psi_{\text{ISM}}$ = 0.11 and $\Psi_{\text{wind}}$ = 0.14, and the distribution widths in log space are $\sigma_{\text{$\Psi$,ISM}} = 0.32$ and $\sigma_{\text{$\Psi$,Wind}} = 0.28$. 

We test for a possible sample bias or correlation between $\Psi$ and the peak flux; see Figure~\ref{fig:psi_fp}. We find a Spearman's rank correlation coefficient of -0.2 or -0.07 between the two parameters, with a p value of 0.08 or 0.6, for the ISM and Wind cases, respectively. This implies that there is no significant relationship between $\Psi$ and the peak flux. We note that our additional Sample 2 has on average higher peak flux values than Sample 1 from \citet{Beniamini_2017}. This is due to the fact that we selected those GRBs for which a peak could be accurately determined from well-sampled light curves or SEDs, while the light curves from \citet{Chandra_Frail_2012} were not necessarily sampled as well.

\subsection{Results for $\chi$}

\begin{figure}
    \centering
    \includegraphics[trim=25 10 25 50, clip, width=\columnwidth]{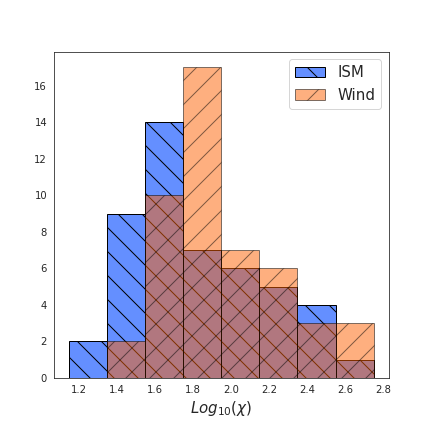}
    \caption{Histogram in log space of $\chi$ at 1~day for our entire GRB sample, for the homogeneous (ISM; blue) and stellar wind (Wind; orange) cases.}
    \label{fig:chi_histogram}
\end{figure}

\begin{figure}
    \centering
    \includegraphics[trim=0 10 25 50, clip, width=\columnwidth]{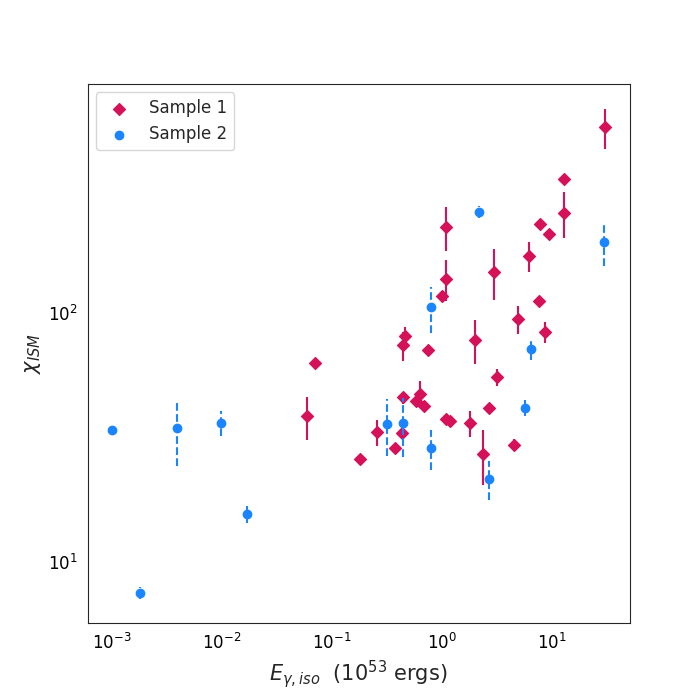}
    \includegraphics[trim=0 10 25 50, clip, width=\columnwidth]{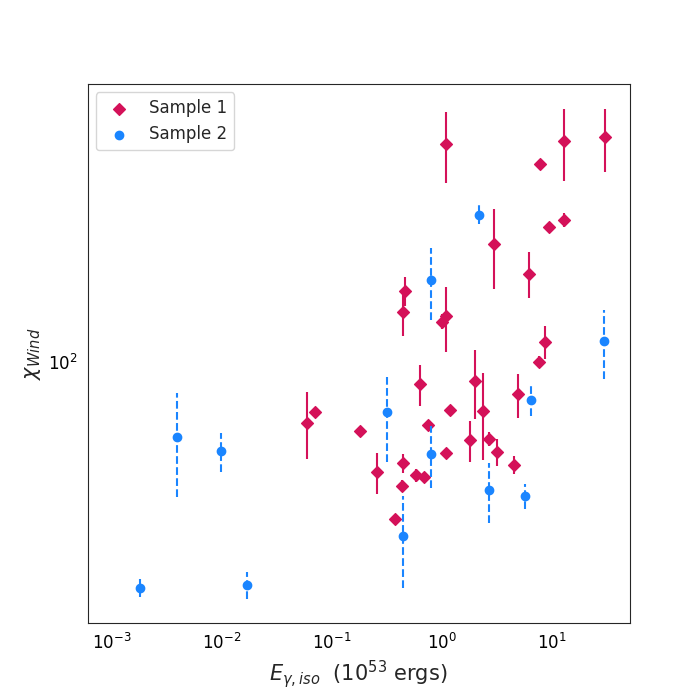}
    \caption{$\chi$ at 1 day versus $E_{\gamma,\text{iso},53}$ for a homogeneous (top panel) and stellar wind (bottom panel) medium, for the two samples presented here.}
    \label{fig:chi_eiso}
\end{figure}

Figure~\ref{fig:chi_histogram} shows a histogram of $\chi$ at 1 day for the ISM (blue) and Wind (orange) ambient medium cases, for all GRBs listed in Table~\ref{tab:psi_chi}. We find that the weighted averages are $\chi_{\text{ISM}}$ = 28 and $\chi_{\text{wind}}$ = 63, and the distribution widths in log space are $\sigma_{\text{$\chi$,ISM}} = 0.36$ and $\sigma_{\text{$\chi$,Wind}} = 0.29$, similar to the distribution widths for $\Psi$. We note that the $\chi$ values and weighted averages will change as a function of time following the time dependence in Equations~\ref{chi_ism} and \ref{chi_wind}, but the widths of the distributions in log space will remain constant.

We examined possible correlations between $\chi$ and any of the observables. We only find a marginal correlation between $\chi$ and $E_{\gamma,\text{iso},53}$, shown in Figure~\ref{fig:chi_eiso}. The Spearman's rank correlation coefficient between these two parameters is 0.5 or 0.4, with a p value of $1\cdot 10^{-5}$ or $4\cdot 10^{-4}$, for the ISM and Wind cases, respectively. This marginal correlation seems to be driven by a couple of GRBs with relatively low and high values for $\chi$ and $E_{\gamma,\text{iso},53}$. Given the marginal significance of the correlation and the strong dependence on a few data points, we do not discuss this possible correlation further.

\section{Discussion} \label{discuss}

GRB afterglows uniquely enable us to study the evolution of the synchrotron spectral peak frequency $\nu_m$, due to the relatively high $\gamma_m$ compared to other jet sources, such as active galactic nuclei and X-ray binaries \citep[e.g.,][]{Jorstad_2001,MillerJones_2006}. In those other cases of relativistic jet emission,  the Lorentz factor of the shocks that are accelerating electrons are too low for $\nu_m$ to be significantly above the synchrotron self-absorption frequency $\nu_a$, which is typically associated with the SED peak in those sources. In particular, $\nu_m$ is proportional to the bulk Lorentz factor of the blast wave to the fourth power \citep{Sari_1998}, which has a large impact for GRBs that have Lorentz factors of tens to hundreds at early times \citep{Rees_1992}. Being able to measure $\nu_m$ and the associated peak flux allows for constraints in GRBs on $\gamma_m$, $\epsilon_e$ and $\xi_e$, as discussed below.

\subsection{Constraints on $\gamma_m$, $\epsilon_e$ and $\xi_e$}

\begin{figure*}
    \centering
    \includegraphics[trim=60 0 90 0, clip, width=0.9\textwidth]{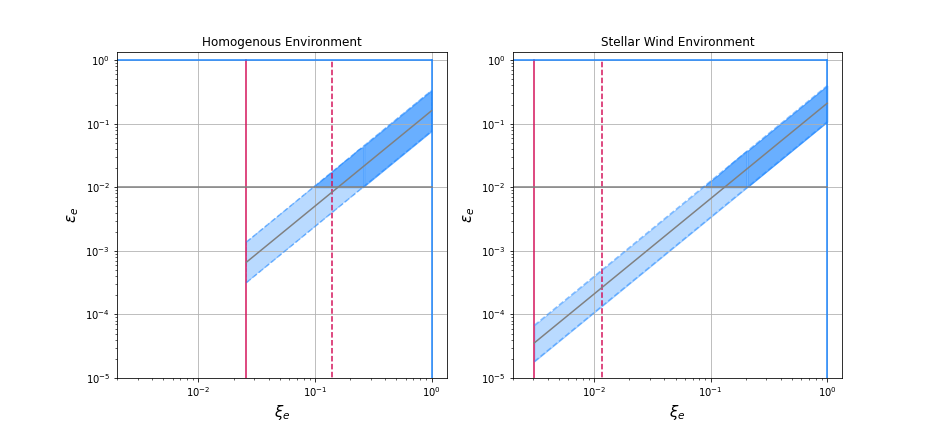}
    \caption{Representation of the lower and upper limits for $\xi$ and $\epsilon_e$, based on $\gamma_m\gtrsim2$ taken at a time of 50 days after the burst (vertical red lines) and $\epsilon_\gamma\gtrsim0.01$ (horizontal grey lines). Solid red lines represent the lower limit of $\xi$ for $\gamma_m\gtrsim2$, and dashed red lines are for $\xi$ plus one standard deviation of $\xi$.}
    \label{fig:limits}
\end{figure*}

As laid out in Section~\ref{sec:method}, $\Psi$ is a proxy for $\epsilon_e~\xi_e ^{-3/2}$, and $\chi$ at a given time is a proxy for $\gamma_{m}~\xi_e^{-1/2}$. For further calculations, we assume values of $\epsilon_\gamma=0.15$ \citep{Beniamini_2015}, $p=2.3$ \citep[e.g.,][]{Ryan_2015}, $n_0=1$ and $A_*=1$ \citep[e.g.,][]{Granot_2014}, with the note that while some of these physical parameters may have relatively broad distributions, $\Psi$ and $\chi$ have weak dependencies on those physical parameters (see also discussion below). From the weighted averages of $\Psi$ and $\chi$, we calculate the most likely values of the following combinations of physical parameters:
\begin{itemize}
    \item $\epsilon_e~\xi_e ^{-3/2} \sim 0.16$ for ISM and $0.21$ for Wind
    \item $\gamma_{m}~\xi_e^{-1/2} \sim 53~t_d^{-3/8}$ for ISM and $96~t_d^{-1/4}$ for Wind
\end{itemize}
One could translate these numbers to values for $\epsilon_e$ and $\gamma_{m}$ directly by making assumptions on $\xi_e$. For instance, the most likely value of $\epsilon_e$ is $0.16-0.21$ for $\xi_e=1$, decreasing to $(5.1-6.6)\cdot 10^{-3}$ for $\xi_e=0.1$ and $(1.6-2.1)\cdot 10^{-4}$ for $\xi_e=0.01$. However, there are lower limits one can set on $\xi_e$ based on the constraints on $\gamma_{m}~\xi_e^{-1/2}$. The peak frequency $\nu_m$ has been identified as such in some well-observed GRB afterglows, such as GRB~970508 and GRB~030329, out to $\sim50$~days \citep[e.g.][]{Frail_2000,Frail_2005,vanderHorst_2005,Granot_2014}. At those times, the GRB blast wave was still relativistic, and we put a conservative lower limit of $\gamma_{m}\gtrsim2$ at $50$~days. This implies lower limits on $\xi_e$ of $\gtrsim 2.5\cdot 10^{-2}$ or $\gtrsim 3.1\cdot 10^{-3}$ for ISM or Wind, respectively.

Given the limits on $\xi_e$, we calculate ranges for $\epsilon_e$ of $6.6\cdot 10^{-4}\lesssim\epsilon_e\lesssim 0.16$ for a homogeneous medium and $3.6\cdot 10^{-5}\lesssim\epsilon_e\lesssim 0.21$ for a stellar wind medium. 

In Figure~\ref{fig:limits}, we show a representation of the limits (solid vertical red lines) on, and dependence (solid diagonal grey lines) of, $\xi_e$ and $\epsilon_e$, for our conservative limit on $\gamma_m$. This figure includes uncertainty regions, bound by diagonal dashed lines, based on the standard deviation in log-space of the $\Psi$ distributions. It also shows the lower limits on $\xi_e$, and therefore on $\epsilon_e$, for taking into account the standard deviation in log-space of the $\chi$ distributions (dashed vertical red lines). In that case, the allowed ranges for both $\xi_e$ and $\epsilon_e$ are significantly smaller. We also note that while $\epsilon_e$ and $\xi_e$ are allowed to both be small, this would have observable consequences: having a low fraction of electrons accelerated into a power-law distribution means that there is a significant thermal electron population, which could be observed at radio wavelengths \citep{Ressler_2017,Warren_2018}. However, since a clear signature of such a thermal electron population has not been observed, $\xi_e$ is likely not much lower than $\sim0.1$, which means that $\epsilon_e$ should have a lower limit of almost $0.01$. As we show in the following sub-section, energy budget considerations constrain $\epsilon_e$ and $\xi_e$ even further, and exclude $\xi_e$ values that would lead to a significant thermal electron population.

\subsubsection{Constraints from $\epsilon_\gamma$}

\citet{Nava_2014} and \citet{Beniamini_2015} put constraints on $\epsilon_\gamma$ and $\epsilon_e$ based on modeling the X-ray and high-energy gamma-ray light curves. They used the prompt gamma-ray fluence and the afterglow light curve as measured by the \emph{Fermi Gamma-ray Space Telescope} (\emph{Fermi}) Large Area Telescope (LAT), and the fact that the latter is in a spectral regime of fast cooling (i.e., $\nu_m>\nu_c$) and without any synchrotron self-Compton cooling effects. They used scalings for the afterglow flux as a function of physical parameters \citep{Granot_2002}, with $\xi_e=1$, to show that $\epsilon_\gamma$ and $\epsilon_e$ have narrow distributions, for $\epsilon_e$ with similar distribution widths as those that we have derived from radio peaks. They found a value of $\epsilon_\gamma\simeq0.15$ for $\epsilon_e\simeq0.1$, and a scaling between these two parameters. In this spectral regime, the scaling of the afterglow flux with almost all the physical parameters is identical between a homogeneous and stellar wind medium:
\begin{equation} \label{LATflux}
\begin{split}
    F_{\nu} &= f(p,\nu,t,z,d_L)~\epsilon_e^{p-1}~\xi_e^{-(p-2)}~E_{iso}^{(p+2)/4}~\epsilon_B^{(p-2)/4}\\
    &= f(p,\nu,t,z,d_L,E_{\gamma,iso})~\epsilon_e^{p-1}~\xi_e^{-(p-2)}~\left(\frac{1-\epsilon_{\gamma}}{\epsilon_{\gamma}}\right)^{(p+2)/4}~\epsilon_B^{(p-2)/4}
\end{split}
\end{equation}
in which $f$ is a function of observables and $p$.

For a large range of $p$ values, the afterglow flux has a very weak dependence on $\epsilon_B$ and also a fairly weak dependence on $\xi_e$. It does have a strong dependence on $\epsilon_e$ and $\epsilon_\gamma$, roughly linear in both. Since the afterglow flux measured in the \emph{Fermi} LAT is narrowly distributed \citep{Nava_2014}, and the flux strongly depends on only $\epsilon_e$ and $\epsilon_\gamma$, any change in one of the parameters leads to a change in the other as well. For instance, $\epsilon_e=10^{-2}$ corresponds to $\epsilon_\gamma\simeq10^{-2}$, and $\epsilon_e=10^{-3}$ to $\epsilon_\gamma\simeq7\cdot10^{-4}$. The latter value for $\epsilon_\gamma$ is unphysical, because it requires an extremely large kinetic energy in the GRB blast wave. We put a conservative lower limit of $\epsilon_\gamma\gtrsim0.01$: for an observed gamma-ray energy of $E_{\gamma,iso}\simeq10^{54}$~erg, a gamma-ray efficiency of 0.01 implies $E_{iso}\simeq10^{56}$~erg and a true, collimation-corrected energy of $\sim10^{53-54}$~erg, which is well above the energetics that can be provided by GRB progenitor and central engine models \citep{Beniamini_2017b}. Therefore, we use this lower limit on $\epsilon_\gamma$ to set a lower limit on $\epsilon_e$ of 0.01. This is indicated in Figure~\ref{fig:limits} as a horizontal grey line, and corresponds to a lower limit of $\xi_e\gtrsim 0.1$.

Given the constraints on $\xi_e$, ranging from $\sim 0.1$ to 1, we put constraints on the range of most likely values for $\gamma_m$ at a given observer time: $(16-53) t_d^{-3/8}$ for ISM and $(30-96) t_d^{-1/4}$ for the Wind case, using the fact that $\gamma_m$ is proportional to $\xi_e^{1/2}$ (see Equations \ref{chi_ism} and \ref{chi_wind}).

\subsection{Comparison to Broadband Modeling}

The methodology presented in \citet{Beniamini_2017} and this paper can be used as a diagnostic tool for broadband modeling studies. Since we focus on only radio light curves and SEDs, the amount of information is limited compared to full-blown modeling across the electromagnetic spectrum. However, the constraints on physical parameters derived from radio afterglow peaks should be consistent with those from broadband modeling. \citet{Granot_2014} compiled broadband modeling studies by various authors using different methodologies, finding a range of $\epsilon_e$ between $4\cdot 10^{-3}$ and $0.3$ (all for $\xi_e=1$). This is a significantly broader range than the widths of our distributions for $\Psi$ or $\epsilon_e$, but this is likely due to the wide variety of methodologies used. 

\citet{Aksulu_2022} studied 22 well-sampled long GRB afterglows to obtain estimates for the physical parameters, using a state-of-the-art, simulations-based modeling methodology. The standard deviation of their distributions for $\epsilon_e$ are 0.24 for ISM and 0.28 for Wind, close to the widths of our distributions for $\Psi$. They find $\epsilon_e$ values between 0.1 and 1, with mean values of 0.34 for a homogeneous medium and 0.28 for a stellar wind medium, which are higher than our findings. We note that a larger density by a factor of a few (Wind) up to $\sim 10$ (ISM) would result in consistent $\epsilon_e$ values; and \citet{Aksulu_2022} indeed find a distribution of densities with a peak that is significantly larger than 1 for both $n_0$ and $A_*$. 

The findings presented in this paper, and recent broadband modeling studies \citep[such as][]{Aksulu_2022} reporting similar distribution widths for, and values of, $\epsilon_e$, shows the validity of using radio peaks as diagnostic tools for broadband modeling. This also means that the distribution widths and values for $\epsilon_e$, $\gamma_m$ and $\xi_e$ presented in this paper can be used for direct comparison with expectations from electron acceleration simulations for relativistic shocks \citep[e.g.,][]{Sironi_2011,Bykov_2012,Marcowith_2016}. However, this requires careful consideration of the effects of all physical parameters on the narrowness of the derived distributions.

\subsection{Universality of physical parameters}

The aforementioned constraints on physical parameters are assuming that the other physical parameters at play are universal among GRB afterglows. Changing $p$ from $2.3$ to $2.1$ or $3$, following distributions presented in the literature \citep[e.g.][]{Starling_2008,Ryan_2015}, does not have a significant effect on the value of $\chi$, but does change $\Psi$ by a factor of $\sim 2.7$ at most, i.e. $\sim 0.4$ in log-space. On the other hand, varying $\epsilon_\gamma$ does not have a significant effect on $\Psi$ but a small effect on $\chi$: changing $\epsilon_\gamma$ by a factor 3, which is quite extreme compared to current constraints \citep{Nava_2014,Beniamini_2015}, results in a factor $\sim 1.6$ change in $\chi$, i.e. $\sim 0.2$ in log-space. 

Varying the density has an effect on both $\Psi$ and $\chi$. The magnitude of variations is quite uncertain, but broadband modeling studies suggest about one order of magnitude larger or smaller \citep{Granot_2014,Aksulu_2022}. This leads to changes in $\Psi$ of factors 1.9 (ISM) or 3.2 (Wind), i.e. 0.27 (ISM) or 0.5 (Wind) in log-space; and to changes in $\chi$ of factors 1.3 (ISM) or 1.8 (Wind), i.e. 0.12 (ISM) or 0.25 (Wind) in log-space.

These variations in $\Psi$ and $\chi$ should be compared to their distribution widths presented in Section~\ref{results}: 0.32 (ISM) and 0.28 (Wind) for $\Psi$, and $0.36$ (ISM) and 0.29 (Wind) for $\chi$. One could conclude from the considerations above that these distribution widths can be explained largely or even completely by variations in the ambient medium density, and allow for (almost) universal values for the physical parameters related to the electron acceleration process in GRB blast waves. However, it is important to keep in mind possible biases in the sample selection that may lead to such conclusions. While there is no significant correlation between $\Psi$ and the radio brightness (see Figure~\ref{fig:psi_fp}), the GRBs in our sample have been selected because a radio peak could be detected. \citet{Chandra_Frail_2012} showed that their sample of radio afterglows was sensitivity limited, and the added sample presented in this paper is biased towards well-sampled radio afterglows. It is possible that a larger range of radio peak fluxes would lead to a broader range of microphysical parameters as well. The peak flux does not depend on $\epsilon_e$, but is proportional to $\xi_e$, so it is possible that GRBs with lower peak fluxes have lower $\xi_e$, and also lower $\epsilon_e$, values.  Addressing this possible bias will require more dedicated observations with current telescopes and deeper observations with new-generation radio observatories such as the Square Kilometer Array \citep[SKA; e.g.,][]{Burlon_2015} and the next-generation Very Large Array \citep[ngVLA; e.g.,][]{McKinnon_2019}.

\subsection{Evolution of Microphysical Parameters} \label{evolution}

\begin{figure} 
    \centering
    \includegraphics[width=\columnwidth]{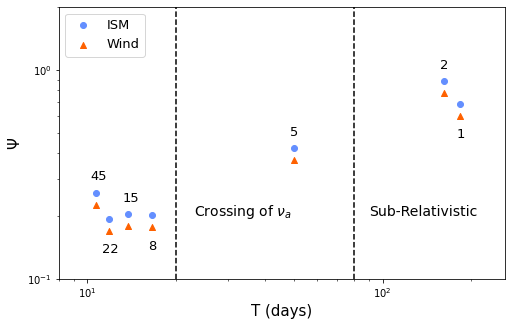}
    \caption{$\Psi$ for GRB 030329 from light curves at different frequencies (indicated in figure, in GHz), for homogeneous (ISM; blue) and stellar wind (Wind; orange) cases, and with vertical lines indicating $\nu_m$ crossing $\nu_a$ and the blast wave becoming sub-relativistic.}
    \label{fig:psi_evolution}
\end{figure}

Up to this point we have assumed that the microphysical parameters are constant as a function of time. Broadband modeling studies \citep[e.g.,][]{Filgas_2011,vanderHorst_2014} and particle acceleration simulations \citep[e.g.,][]{Lemoine_2015} indicate that $\epsilon_e$ may be time variable, although the variability could be mostly in the magnetic field rather than the electron acceleration. Given that the $\epsilon_e$ distribution is narrow, it is unlikely that there is a strong temporal evolution of $\epsilon_e$, unless it is always similar around the typical radio peak observations.

One way to investigate the possible temporal evolution of $\epsilon_e$ is to calculate this parameter from SEDs at different times or light curves at different radio frequencies; and covering a large time span, so variability can be measured or strongly constrained. A good case study for this is GRB~030329, which was observed across several radio frequencies, with well-determined peaks in at least 7 of them. Figure~\ref{fig:psi_evolution} shows the values of $\Psi$ as a function of time, based on the light curves presented in \citet{vanderHorst_2008}, and assuming that the light curve peak is the passage of $\nu_m$ across the observing band. The first four light curve peaks result in similar values for $\Psi$, but at later times the value of $\Psi$ deviate significantly. This is not surprising, given that broadband modeling has shown that $\nu_m$ crosses $\nu_a$ at $\sim 20$~days, making $\nu_a$ the peak of the spectrum after that time; and the blast wave becomes sub-relativistic and eventually non-relativistic at some point, violating the assumptions for our calculations of $\Psi$. Since the derived $\Psi$ values before $20$~days do not evolve, there does not seem to be any temporal evolution of $\epsilon_e$ in this particular GRB afterglow. This will need to be further tested with other GRBs that have multiple radio peaks spanning a large time range.

\begin{table*}
\caption{Values of $\Psi$ and $\chi$ calculated for the entire sample of GRB afterglows. Italicized lines indicate GRBs with a second peak that are not included in sample calculations (and only included for comparison purposes).}
\begin{tabular}{lllll}
\hline \hline
Name    & $\Psi_{\text{ISM}}$ & $\Psi_{\text{Wind}}$ & $\chi_{\text{ISM}}$ & $\chi_{\text{Wind}}$ \\ \hline
190829A & 0.25 $\pm$ 0.08   & 0.53 $\pm$ 0.16    & 35 $\pm$ 10       & 64 $\pm$ 19        \\
171010A & 0.78 $\pm$ 0.04   & 0.38 $\pm$ 0.02    & 256 $\pm$ 14      & 238 $\pm$ 13       \\
161219B & 0.059 $\pm$ 0.003  & 0.26 $\pm$ 0.01    & 8 $\pm$ 1         & 26 $\pm$ 1         \\
\emph{160625B} & \emph{0.20 $\pm$ 0.09}    & \emph{0.06 $\pm$ 0.02}    & \emph{116 $\pm$ 48}      & 7\emph{8 $\pm$ 32}        \\
160625B & 0.34 $\pm$ 0.07   & 0.08 $\pm$ 0.02    & 193 $\pm$ 39      & 113 $\pm$ 23       \\
160509A & 0.093 $\pm$ 0.007  & 0.050 $\pm$ 0.004   & 42 $\pm$ 3        & 45 $\pm$ 3         \\
\emph{160509A} & \emph{0.076 $\pm$ 0.005}  & \emph{0.047 $\pm$ 0.003}   & \emph{34 $\pm$ 2}        & \emph{42 $\pm$ 3}         \\
150413A & 0.45 $\pm$ 0.04   & 0.22 $\pm$ 0.02    & 195 $\pm$ 18      & 198 $\pm$ 18       \\
141121A & 0.29 $\pm$ 0.06   & 0.29 $\pm$ 0.06    & 106 $\pm$ 22      & 161 $\pm$ 34       \\
140713A & 0.076 $\pm$ 0.006  & 0.13 $\pm$ 0.01    & 16 $\pm$ 1        & 27 $\pm$ 2         \\
140311A & 0.037 $\pm$ 0.007  & 0.050 $\pm$ 0.009   & 22 $\pm$ 4        & 47 $\pm$ 8         \\
120326A & 0.11 $\pm$ 0.03   & 0.16 $\pm$ 0.04    & 36 $\pm$ 9        & 74 $\pm$ 19        \\
\emph{120326A} & \emph{0.09 $\pm$ 0.02}   & \emph{0.14 $\pm$ 0.02}    & \emph{31 $\pm$ 5}        & \emph{65 $\pm$ 10}        \\
111215A & 0.10 $\pm$ 0.03    & 0.07 $\pm$ 0.02    & 37 $\pm$ 10       & 37 $\pm$ 10        \\
100901A & 0.08 $\pm$ 0.01   & 0.10 $\pm$ 0.02     & 29 $\pm$ 5        & 58 $\pm$ 11        \\
100814A & 0.127 $\pm$ 0.005  & 0.099 $\pm$ 0.004    & 44 $\pm$ 2        & 51 $\pm$ 2         \\
\emph{100418A} & \emph{0.20 $\pm$ 0.12}    & \emph{0.35 $\pm$ 0.20}    & \emph{36 $\pm$ 21}       & \emph{59 $\pm$ 33}        \\
100418A & 0.20 $\pm$ 0.02    & 0.35 $\pm$ 0.04    & 36 $\pm$ 4        & 59 $\pm$ 7         \\
100414A & 0.234 $\pm$ 0.009  & 0.101 $\pm$ 0.004    & 112 $\pm$ 4       & 100 $\pm$ 4        \\
091020  & 0.131 $\pm$ 0.007  & 0.110 $\pm$ 0.006   & 46 $\pm$ 3        & 55 $\pm$ 3         \\
090902B & 0.9 $\pm$ 0.1    & 0.26 $\pm$ 0.05    & 561 $\pm$ 104     & 375 $\pm$ 69       \\
090715B & 0.05 $\pm$ 0.01   & 0.09 $\pm$ 0.02    & 27 $\pm$ 7        & 75 $\pm$ 19        \\
090424  & 0.26 $\pm$ 0.04   & 0.31 $\pm$ 0.04    & 74 $\pm$ 10       & 134 $\pm$ 18       \\
090423  & 0.22 $\pm$ 0.04   & 0.15 $\pm$ 0.03    & 138 $\pm$ 26      & 131 $\pm$ 25       \\
090328  & 0.36 $\pm$ 0.02   & 0.23 $\pm$ 0.01    & 117 $\pm$ 5       & 127 $\pm$ 6        \\
090313  & 0.052 $\pm$ 0.003  & 0.054 $\pm$ 0.003   & 30 $\pm$ 2        & 54 $\pm$ 3         \\
071010B & 0.12 $\pm$ 0.01   & 0.13 $\pm$ 0.02    & 33 $\pm$ 4        & 52 $\pm$ 6         \\
071003  & 0.13 $\pm$ 0.01   & 0.072 $\pm$ 0.006   & 55 $\pm$ 4        & 59 $\pm$ 5         \\
070125  & 0.42 $\pm$ 0.01   & 0.209 $\pm$ 0.005   & 209 $\pm$ 5       & 221 $\pm$ 5        \\
051022  & 0.41 $\pm$ 0.06   & 0.19 $\pm$ 0.03    & 170 $\pm$ 23      & 168 $\pm$ 23       \\
050904  & 0.45 $\pm$ 0.02   & 0.155 $\pm$ 0.007   & 346 $\pm$ 15      & 231 $\pm$ 10       \\
050820A & 0.17 $\pm$ 0.03   & 0.11 $\pm$ 0.02    & 78 $\pm$ 16       & 89 $\pm$ 18        \\
050603  & 0.18 $\pm$ 0.02   & 0.083 $\pm$ 0.011   & 95 $\pm$ 12       & 82 $\pm$ 11        \\
031203  & 0.291 $\pm$ 0.007   & 2.80 $\pm$ 0.08     & 152 $\pm$ 4       & 243 $\pm$ 7        \\
030329  & 0.113 $\pm$ 0.001  & 0.206 $\pm$ 0.001   & 26 $\pm$ 0        & 66 $\pm$ 1         \\
030226  & 0.089 $\pm$ 0.001  & 0.115 $\pm$ 0.002   & 37 $\pm$ 1        & 75 $\pm$ 1         \\
021004  & 0.077 $\pm$ 0.002  & 0.078 $\pm$ 0.002   & 29 $\pm$ 1        & 40 $\pm$ 1         \\
020819B & 0.53 $\pm$ 0.05   & 0.57 $\pm$ 0.05    & 117 $\pm$ 11      & 157 $\pm$ 14       \\
020405  & 0.7 $\pm$ 0.1   & 0.7 $\pm$ 0.1    & 222 $\pm$ 45      & 360 $\pm$ 73       \\
011211  & 0.12 $\pm$ 0.02   & 0.16 $\pm$ 0.02    & 47 $\pm$ 6        & 88 $\pm$ 11        \\
011121  & 0.30 $\pm$ 0.03    & 0.36 $\pm$ 0.03    & 81 $\pm$ 7        & 151 $\pm$ 13       \\
010222  & 0.5 $\pm$ 0.1   & 0.32 $\pm$ 0.07    & 254 $\pm$ 53      & 367 $\pm$ 76       \\
000926  & 0.090 $\pm$ 0.004  & 0.079 $\pm$ 0.003   & 41 $\pm$ 2        & 63 $\pm$ 3         \\
000911  & 0.18 $\pm$ 0.02   & 0.11 $\pm$ 0.01    & 84 $\pm$ 8        & 112 $\pm$ 11       \\
000418  & 0.209 $\pm$ 0.003  & 0.128 $\pm$ 0.002   & 71 $\pm$ 1        & 69 $\pm$ 1         \\
000301C & 0.091 $\pm$ 0.003  & 0.094 $\pm$ 0.003   & 33 $\pm$ 1        & 48 $\pm$ 2         \\
991208  & 0.114 $\pm$ 0.001  & 0.105 $\pm$ 0.001   & 38 $\pm$ 0        & 58 $\pm$ 1         \\
990510  & 0.09 $\pm$ 0.01   & 0.09 $\pm$ 0.01    & 36 $\pm$ 4        & 63 $\pm$ 8         \\
981226  & 0.16 $\pm$ 0.03   & 0.25 $\pm$ 0.05    & 39 $\pm$ 8        & 70 $\pm$ 14        \\
980703  & 0.129 $\pm$ 0.003  & 0.098 $\pm$ 0.002    & 42 $\pm$ 1        & 51 $\pm$ 1         \\
970826  & 0.37 $\pm$ 0.09   & 0.27 $\pm$ 0.06    & 147 $\pm$ 34      & 199 $\pm$ 46       \\
970508  & 0.263 $\pm$ 0.003  & 0.260 $\pm$ 0.003   & 63 $\pm$ 1        & 74 $\pm$ 1        
\end{tabular}
\label{tab:psi_chi}
\end{table*}

\section{Conclusions} \label{conclusions}

We have presented a methodology to constrain physical parameters of electron acceleration in GRB blast waves based on radio afterglow peaks, building on the methodology presented in \citet{Beniamini_2017}. Our methodology provides independent constraints on $\gamma_m$, $\epsilon_e$ and $\xi_e$, and we have shown that it can be used to complement and corroborate modeling efforts across the electromagnetic spectrum. We have expanded the sample to 49 GRB afterglows and confirmed the initial findings in \citet{Beniamini_2017}. We have also shown that one can use peaks of SEDs, as well as light curves, to constrain the aforementioned physical parameters.

We derived distributions for $\Psi$, which is a proxy for $\epsilon_e~\xi_e ^{-3/2}$, and $\chi$, which is a proxy for $\gamma_{m}~\xi_e^{-1/2}$ at a given time. For conservative lower limits on $\gamma_{m}$, we derived allowed parameter ranges of $2.5\cdot10^{-2}\lesssim\xi_e\lesssim 1$ and $6.6\cdot 10^{-4}\lesssim\epsilon_e\lesssim 0.16$ for a homogeneous medium, and $3.1\cdot10^{-3}\lesssim\xi_e\lesssim 1$ and $3.6\cdot 10^{-5}\lesssim\epsilon_e\lesssim 0.21$ for a stellar wind medium. We increase the lower limits even further by adopting that the energy efficiency of the prompt gamma-ray emission, $\epsilon_\gamma$, should be at least $0.01$ to avoid unphysical blast wave energetics. This leads to lower limits of $\xi_e\gtrsim 0.1$ and $\epsilon_e\gtrsim 0.01$, which is consistent with the lack of a clear signature of a thermal electron population in radio observations. This means that we have constrained the values of $\epsilon_e$ and $\xi_e$ to within about one order of magnitude. We note that the $\epsilon_e$ values for $\xi_e=1$ are consistent with those found by state-of-the-art broadband modeling studies. 

The distributions for $\Psi$ and $\chi$ are fairly narrow, with a standard deviation of $\sim 0.3$ in log-space. These distribution widths can potentially be attributed to variety in ambient medium densities and the power-law index of the accelerated electron energy distribution, leading to very narrow distributions in $\epsilon_e$, $\xi_e$ and $\gamma_m$. However, we cannot conclude that there is universality among GRBs when it comes to the microphysical parameters describing electron acceleration, since our sample may be biased towards GRBs with bright radio afterglows. Follow-up studies with radio-dimmer GRBs, either with current or new instrumentation, will probe this universality question further. We also showed how future observations of radio peaks can address possible temporal evolution of $\epsilon_e$ or $\xi_e$.

\section*{Acknowledgements}
The authors would like to thank the anonymous referee for their useful comments that strengthened the manuscript. RAD would like to thank Michael Moss for assistance with Python programming and feedback on this paper; and Sarah Chastain for assistance with with data collection. RAD would also like to thank the George Washington University Undergraduate Research Award and the Physics Department Frances E. Walker Fellowship for financial support to carry out the research presented in this paper. PB was supported by a grant (no. 2020747) from the United States-Israel Binational Science Foundation (BSF), Jerusalem, Israel.

\section*{Data Availability}

The data underlying this article are available in the article and in its online supplementary material. The modeled data sets were derived from sources in the public domain.

\bibliographystyle{mnras}
\bibliography{citations}

\bsp
\label{lastpage}
\end{document}